\documentclass{kapproc} 

\usepackage{procps} 

\usepackage[dvips]{graphicx}

\upperandlowercase

\let\footnote\savefootnote

\kluwerbib 

\def\z{\zeta}

\def\L{\Lambda}

\begin{document}



\articletitle[Scaling of voids]{Scaling of voids in\\ the large scale
distribution of matter}


\author{Jose Gaite}









--------------

\affil{Instituto de
Matem{a}ticas y F{i}sica Fundamental, CSIC, Serrano 123, 28006
Madrid, Spain.}
\email{jgaite@imaff.cfmac.csic.es}


\begin{abstract}
Voids are a prominent feature of the galaxy distribution but their
quantitative study is hindered by the lack of a precise definition of
what constitutes a void. Here we propose a definition of voids
in point distributions that
uses methods of discrete stochastic geometry, in particular,
Delaunay and Voronoi tessellations, and we construct a new void-finder.
We then apply the void-finder to scaling point distributions.
First, we find the voids of pure fractals with a transition to
homogeneity and show that the rank ordering of the voids also scales 
(Zipf's law) and, in addition, shows the transition to homogeneity. 
However, a pure fractal is
arguably not a good model of the galaxy distribution, so we construct
from a cosmological $N$-body simulation a {\em bifractal} mock galaxy sample 
representing two galaxy populations, which we identify as ``wall''
and ``field'' galaxies. The wall galaxy distribution fits
a pure fractal with a transition to homogeneity and, furthermore,
the rank ordering of its voids show a scaling range with
the right slope plus a transition to homogeneity.
\end{abstract}


\section*{Introduction}

The standard statistical analysis of the distribution of galaxies
consists in the derivation of the correlation functions, chiefly, the
two-point correlation function $\xi$.  The large value of this
correlation function ($\xi \gg 1$) on small scales is interpreted as the
tendency of galaxies to cluster strongly. The reverse side of strong
clustering is the appearance of void regions, with essentially no
galaxies. Voids have an intrinsic appeal, since our sensory system is
well suited to perceive patterns, so we immediately perceive void
shapes while looking at a point set. The
observation of voids in the distribution of galaxies has led to the
compilation of void catalogues and to their statistical analysis
(Einasto, Einasto \& Gramann, 1989; Vogeley, Geller \& Huchra, 1991;
El-Ad, Piran \& da Costa, 1997, Mueller et al., 2000; Hoyle \&
Vogeley, 2002). A precise definition of void is however a delicate
matter, and it is a moot point to determine to what extent the
conclusions drawn from statistics of voids depend on the particular
definition adopted.

Analytic models of structure formation such as the adhesion model as
well as numerical simulations show that the distribution of matter
and, in particular, the distribution of galaxies constitutes an
interpenetrating network of clusters and voids, which has been dubbed
``the cosmic web''. This network has a self-similar nature.  On the
other hand, the hierarchical structure of clusters of galaxies and the
analysis of the two-point correlation function provides evidence for a
self-similar {\em fractal} structure (Mandelbrot, 1977, Coleman \&
Pietronero, 1992; Sylos Labini, Montuori \& Pietronero, 1998) or {\em
multifractal} structure (Mart{i}nez et al, 1990; Jones et al,
1992).
It was noted years ago that the cosmic web seems to exhibit a
hierarchy of voids, but the scaling properties of voids in the
distribution of galaxies were hardly studied (as an exception,
see Einasto, Einasto \& Gramann, 1989).

Here, we begin by recalling the properties of voids in fractal
distributions and its application to the galaxy distribution derived
in previous work (Gaite \& Manrubia, 2002).  Self-similarity is the
most obvious property and manifests itself in the scaling of void
sizes, which can be conveniently described by the {\em Zipf law} (a
rank-ordering power law), as shown by Gaite \& Manrubia (2002) using
regular shape voids. We also showed 
how to obtain
the fractal dimension
from this law. 
However, the lists of voids in galaxy
catalogues that we examined fail to show any scaling. We atributed
this to shortcomings of current void searching algorithms, regarding
their definition of void. Therefore,
the analysis of void sizes and their scaling properties is a promising
tool but it cannot be really effective until having a suitable
definition of void. Here we propose a new definition
of void and a new void-finder (using that definition), and explore
its application to the scaling properties of cosmological 
matter distributions.

Existing definitions of void and void-finders differ. We must 
mention: 
(i) the empty sphere method (Einasto, Einasto \& Gramann, 1989);
(ii) the improvement with elliptical regions (Ryden \& Melott, 1996);
(iii) the progressive construction of voids with cubes + rectangular prisms 
(Kaufmann \& Fairall, 1991);
(iv) the related method that uses connected spheres 
(El-Ad \& Piran, 1997);
(v) the method of distance field maxima (Aikio \& Mahonen, 1998);
(vi) the use of the smoothed density field (Shandarin, Sheth \& Sahni, 2004).
For us it is natural to
rely on the methods of discrete stochastic geometry, namely, 
Delaunay and Voronoi tessellations, the use of which in
Cosmology has been pioneered by Rien van de Weygaert and collaborators
(Schaap \& van de Weygaert, 2000).  

We will first describe our void-finder and its geometrical basis.  The
problem of definition of void is similar in two or higher dimensions,
so we will begin with two-dimensional fractal point sets (with
transition to homogeneity), whose voids are easy to visualize. This is
useful since the notion of void arises as an intuitive visual notion.
The conclusions can be extrapolated to three-dimensional fractals.
Next, we show how to apply the method to the galaxy distribution, by
means of mock distributions obtained from cosmological $N$-body
simulations. Finally, we discuss the results.

\section[]{Discrete geometry methods and void finder}

The Delaunay tessellation of a three-dimensional point set consists of
a set of links between points forming simplices such that their respective
circumscribing spheres do not contain any other point of the set. This
tessellation is unique and, moreover, is particularly suited for the
search for voids because each circumscribing sphere is
void. The Delaunay tessellation can be generalized to any dimension and,
in particular, in two dimensions is called the Delaunay triangulation.
There is a dual tessellation formed by the centers of the circumscribing
(hyper)spheres, called the Voronoi tessellation. Each Voronoi cell is
the neighbourhood of a point of the set, in the sense that the points of
the interior of the cell are closer to that point than to any other point
of the set.

The Delaunay and Voronoi tessellations are fundamental constructions
associated to a point set. In the search for voids, the defining
property of the Delaunay tessellation is obviously adequate for the
definition of voids: we can consider each simplex as an elementary
void. Then we grow a given elementary void by joining adjacent
simplices if appropriate, according to some criteria. Consequently, we
define a void as set of adjacent in pairs simplices with a boundary
given by the separation criteria.

The natural separation criterium between elementary spherical voids is
given by the fractional overlap being below a predetermined threshold,
as used by Hoyle \& Vogeley (2002). Naturally, our elementary
spherical voids are the circumscribing spheres
corresponding to adjacent simplices. However, this criterium is not
sufficient and we need an additional condition on the relative
diameter of the added simplices to prevent the merging of elementary
voids of very different size, similar to the prescription of El-Ad \&
Piran (1997).
So our procedure relies on preceding void-finders. Its advantage is
that the elementary spheres are given by the Delaunay tessellation.

To begin the search for voids we need to estimate where we may find
the largest one, but we cannot measure the size of a void until it has
been found; so we look for the largest Delaunay simplex. The algorithm
consists of the following steps:

\begin{enumerate} \itemsep 0mm
\item Construct the Delaunay and Voronoi tessellations for the
given point set.
\item Sort the simplices of the Delaunay triangulation by size and select
the largest one to begin to build the first void.
\item Grow the void by adding adjacent simplices (in this process, the
Voronoi tessellation is useful). A simplex is added if the overlap
criterium is met and the ratio of the diameter 
of its touching face to the diameter of the initial 
simplex is above a given value (``link ratio'').  The set of all
simplices found constitutes the void.
\item Iterate by finding the largest simplex among the remaining ones
until they are exhausted.
\end{enumerate}

The algorithm is valid in any dimension but we have only applied it to
two and three dimensional point sets.

\section[]{Properties of voids in fractal distributions}

Our void finding algorithm produces a list of voids roughly ordered by
their size. Actually, one must measure the sizes of voids an reorder
them accordingly. Thus one gets the voids listed by
decreasing size, which is suitable for a study with rank-ordering
techniques, that is, for studying how the size of voids decreases with
rank. If there is a
hierarchy of voids, one must expect a {\em regular} decreasing; in
particular, a power-law decreasing is the Zipf law (that is, a simple
linear decrease in the log-log plot). The interest of this law is that
it does not mark any scale and it is, therefore, associated with scale
invariance (Gaite \& Manrubia, 2002). Concretely, a scaling fractal
set must have a scaling void hierarchy. This is simple to prove in one
dimension, in which the voids are well defined as intervals, but a proof 
in higher dimensions requires before a definition of void, such as 
the one used in our void-finder. 
So we will apply it to scaling fractal sets to test the Zipf law. 

Firstly, let us consider, in one dimension,
how the transition to homogeneity is achieved
and its effect on the Zipf law. A transition to homogeneity
in a one-dimensional random Cantor-like fractal can be achieved by
joining by the ends several realizations of it.  For the sake of the
argument, let us assume that a single realization follows a perfect
Zipf law $\Lambda_R = \L_1 R^{-\z}$. Then we have in 10 copies of the
fractal, say, 10 voids with size $\L_1$ and ranks $R = 1, \ldots, 10$,
10 voids with size $\L_2$ and ranks $R = 11, \ldots, 20$, etc. So the
sizes follow the law $\log\L_{10 n} = \log\L_{1} + \z \log{10} - \z
\log{(10 n)}$, $\L$ being constant between ranks $10 n - 9$ and $10
n$. This is a stepcase with steps of exponentially decreasing width
and linearly descending ends with slope $\z$.  Relaxing the condition
of an initial perfect Zipf law we smooth the steps, so we conclude
that the effect of the transition to homogeneity is the flattening of
the Zipf law for small ranks and that the width of the flattened
portion measures the scale of homogeneity.  

In any dimension, we can generate periodic pure fractals (with
transition to homogeneity) 
with a method based on the theory of fractional
Brownian motion. We must check the fractal 
by computing its number-radius function, which has to be a power law,
$N(r) \propto r^D$ 
(equivalent to $\xi \propto r^{D-3}$ {\em and} $\xi \gg 1$).
This law must hold for a sufficient range of scales between the 
minimum inter-point distance and the scale of homogeneity.


\begin{figure}
\includegraphics[width=7cm]{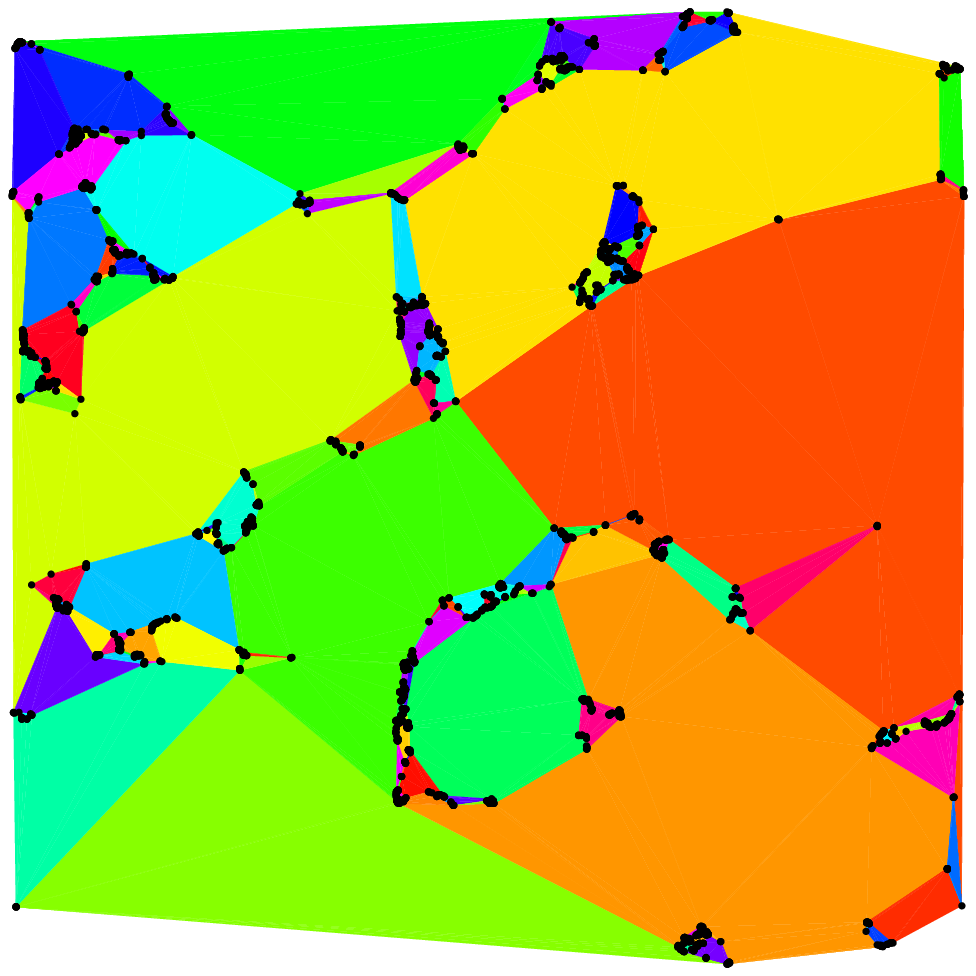}
\narrowcaption{Voids of a random fractal set
with $D= 0.8$ and 3871 points.}
\label{voids}
\end{figure}

\begin{figure}
\includegraphics[width=7.5cm]{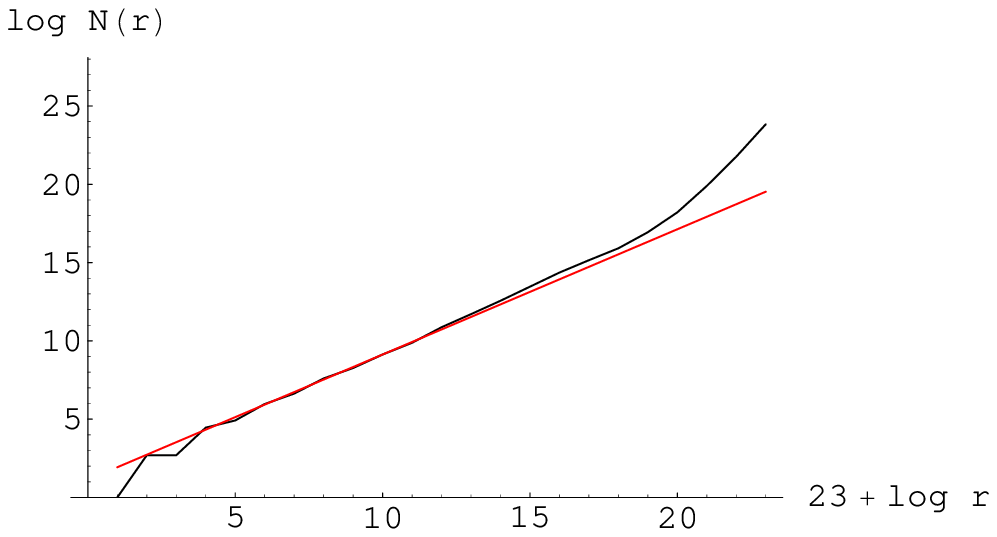}\\
\includegraphics[width=7.5cm]{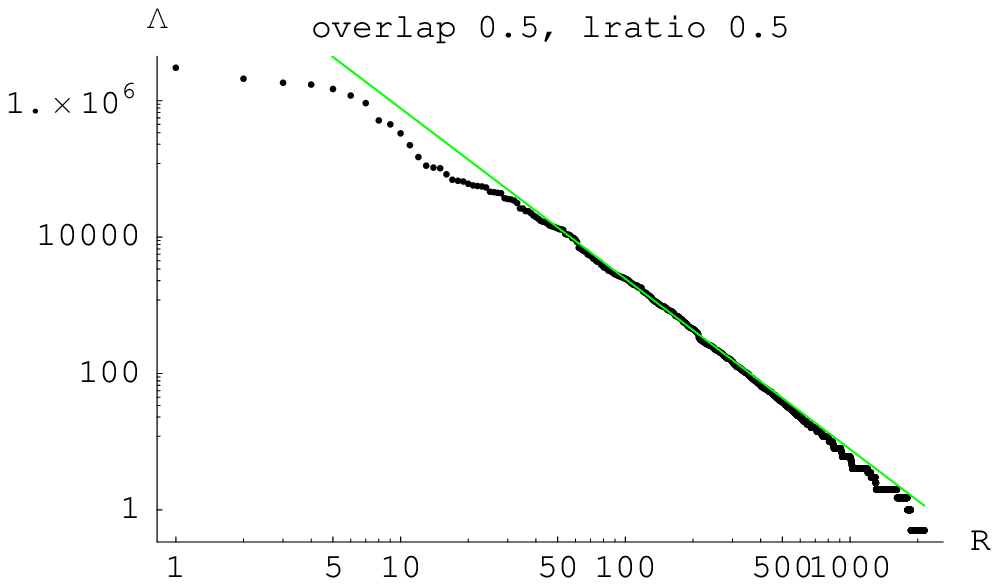}
\vskip -.5cm
\narrowcaption{Number-radius function of the $D=0.8$ fractal point set in
$d=2$ and rank ordering of voids (corresponding to the displayed void-finder
setting).}
\label{2d}
\end{figure}

In two dimensions, we have run our void-finder on various fractal point
sets with different dimensions and various values of the overlap and
link ratio thresholds. The Zipf law obtains, but to have the maximum
scaling range, that is, similar to the scaling range of the number
function, one has to tune the algorithm parameters.  Fig.~\ref{voids}
shows the voids found in a fractal with dimension $D= 0.8$ and 3871
points (for both overlap and link ratio $= 0.5$).  
Fig.~\ref{2d} shows its number-radius function and the Zipf's
law for voids.  We observe that the transition to homogeneity 
has a similar aspect in both plots.
The effect of the
transition to homogeneity in the rank-ordering of voids
is the flattening of the Zipf law for small
ranks and the width of the flattened portion measures the scale
of homogeneity.
So an essential feature of both
graphs is that there is a crossover between two different 
regions. The left flat region in the rank-ordering of voids
corresponds to the transition to homogeneity whereas the
right scaling region corresponds to the
expected scaling with slope $-2/D = -2.5$. In contrast, 
in the graph of the number-radius function, the transition
to homogeneity corresponds to the right region, where the slope
tends to 3.

The rank-ordering of voids and the
corresponding Zipf's law convey no more information than the
number-radius function and, in fact, one needs to tune somewhat the
void-finder parameters to extract the same information. This is due to
the fact that the morphology of voids depends on the type of
fractal. The information on morphology (including features like
{\em lacunarity}) has value of its own but the existence of various
morphologies corresponding to the same scaling dimension poses
difficulties for void-finders, which need to adapt to a particular
morphology.  Obviously, the void-finder parameters must only depend on
relative magnitudes, like the overlap and link ratio which we have used, 
but one has some liberty in their choice.

\section[]{Bifractal distribution: field and wall galaxies}

Before looking for voids in a galaxy sample, it is usual to divide the
sample between wall and field galaxies and remove the latter (El-Ad \&
Piran, 1997).  The rationale for this is that the galaxy distribution
is very inhomogeneous, in the sense that there are very rich clusters
with high luminosity galaxies and there also are faint galaxies which
are weakly clustered. In fact, this idea can be extended
to the full dark matter distribution, in connection with the concept
of {\em bias}: rich galaxy clusters are located in places with very
high density of dark matter (massive haloes) and fainter galaxies in
places with moderate density (normal haloes), whereas the regions of
very low density of dark matter are depleted from galaxies. The
definition and structure of voids have already been examined in this
context (Benson et al, 2003; Gottloeber et al, 2003).

The pure fractals we have been using are not suitable to incorporate
the idea of two different populations of galaxies or, more in general,
the idea of biasing. This is because every particle of a pure fractal
has the same properties in relation with the other particles. This
uniformity characterizes {\em monofractal} distributions, as opposed
to {\em multifractal} distributions, which can be pictured as a
superposition of monofractals with various dimensions (Mart{i}nez et
al, 1990; Jones, Coles \& Mart{i}nez, 1992).

\begin{figure}
\includegraphics[width=7cm]{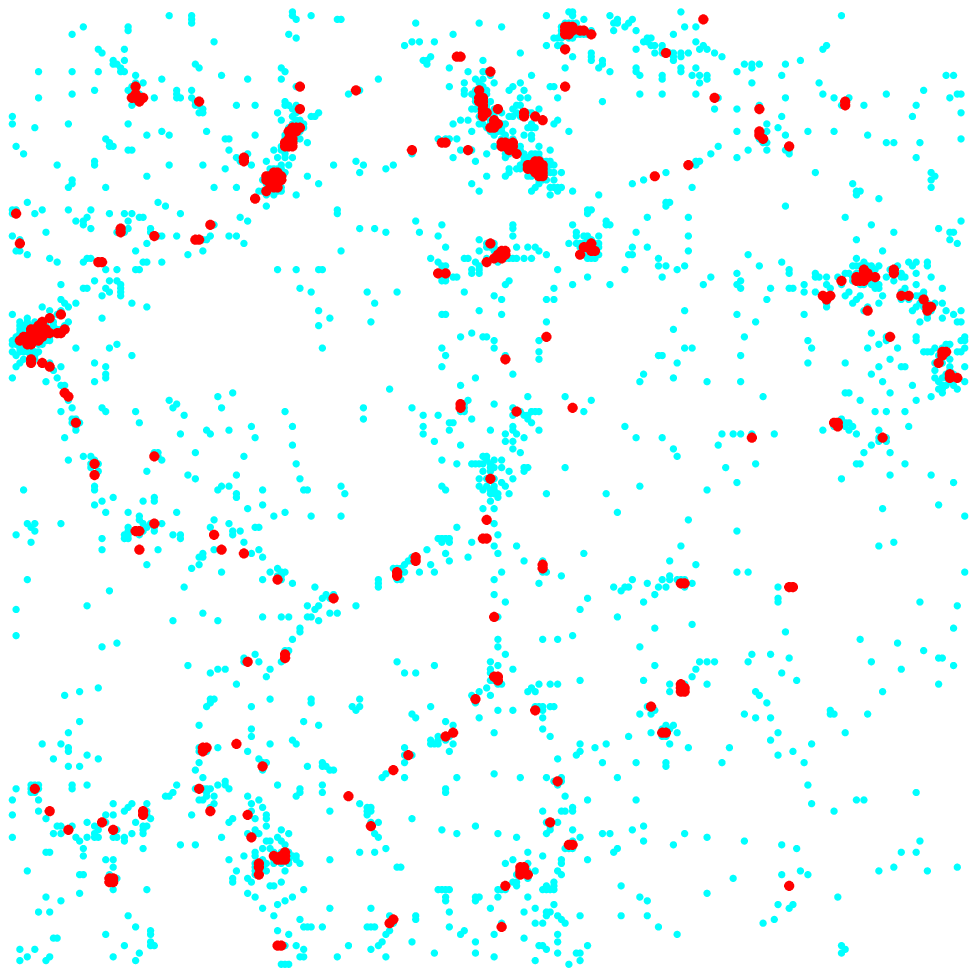}
\narrowcaption{Selection of two populations in a slice: in red the
more clustered population ($M = 3539$) and in blue the less clustered
population ($M = 2255$).}
\label{2pop}
\end{figure}

The general multifractal model of the dark matter distribution assumes
a superposition of monofractals with a continuous range of dimensions
(in principle, running from the lowest to the highest, namely, from
zero to three). Therefore, it would be reasonable to associate to it a
multifractal distribution of galaxies.  This continuous superposition
of monofractals would be impractical for void searching, so we use a
cutoff to
obtain a two-population distribution that we
identify with the wall and field galaxy populations.
A {\em bifractal} distribution has also been favoured
on more fundamental grounds (Balian \& Schaeffer, 1988).

We have applied this bifractal model to the distribution of dark
matter halos in a simulation with the Hydra code, namely, the $z=0$
positions of a simulation with $86^3$ particles (sufficient to test
our method).  We identify dark matter haloes with overdensities (by
means of a spatial window) and we populate haloes with galaxies
assuming linear bias.  To illustrate the appearance of both
populations, we show in Fig.\ \ref{2pop} the result of this process 
on a slice of the Hydra simulation. 
Note that the more clustered population (with
lower dimension) has larger total mass in spite of being sparser.

\begin{figure}
\includegraphics[width=7cm]{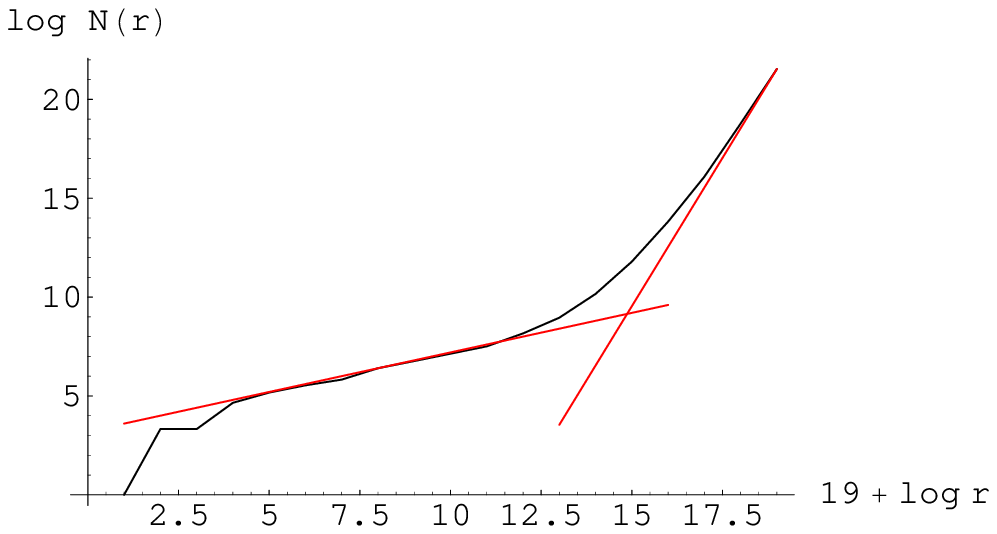}
\vskip -.5cm 
\narrowcaption{Number-radius relation for a selection of
$1\,746$ strongly clustered Hydra ``wall galaxies''.
It has a scaling range, corresponding to a fractal dimension $D = 0.4$,
and a transition to homogeneity.}
\label{dim-H}
\end{figure}

One can split the galaxies in two populations in several ways by
assigning different halo mass cutoffs. We have found convenient to use a
population of $1\,746$ ``wall galaxies'' (corresponding to halos with
total mass $M=22\,738$). In any division in two populations,
for consistency, we
must check that each population approximately corresponds to a
monofractal. This is especially important for the ``wall galaxy''
population. We have calculated the number-radius relation for the
population of $1\,746$ ``wall galaxies'' and indeed found a power law in
a scale range, corresponding to a surprisingly small dimension,
namely, $D = 0.4$.  Furthermore, the transition to homogeneity on the
largest scales is clearly visible (see Fig.\ \ref{dim-H}).

\begin{figure}
\includegraphics[width=6cm]{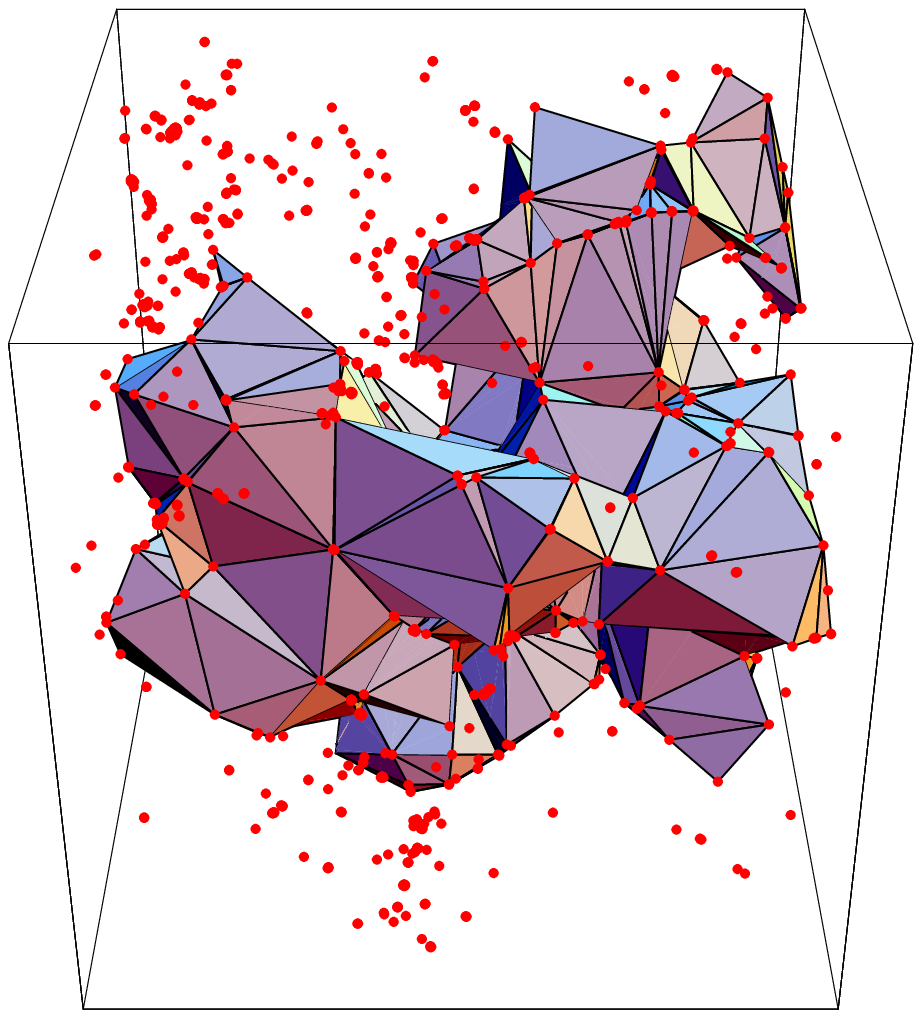}
\narrowcaption{
Voids with 2nd to 6th rank in the distribution of 
the $1\,746$ Hydra ``wall galaxies''.}
\label{voids-H}
\end{figure}

\begin{figure}
\centering{\includegraphics[width=8.5cm]{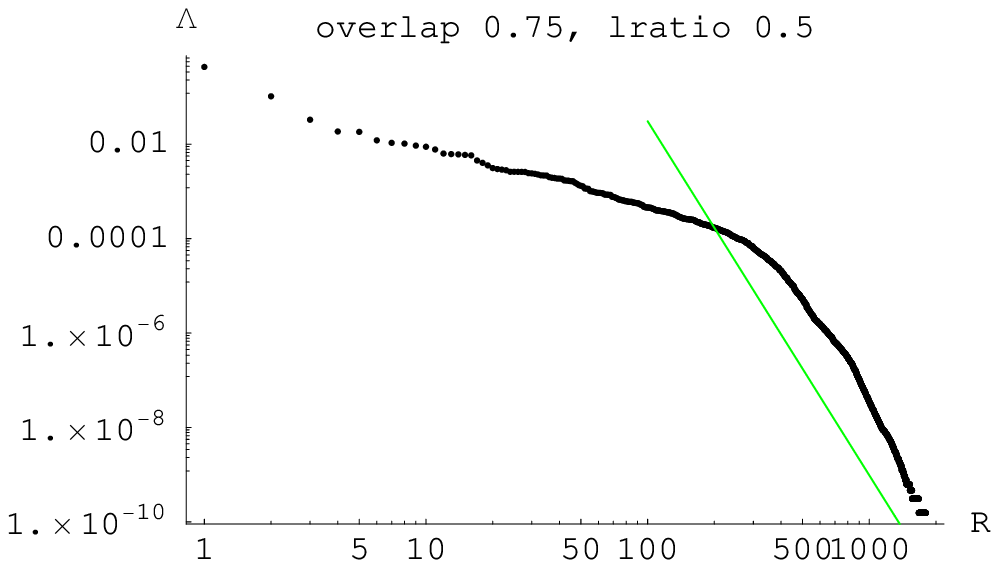}}
\caption{Rank ordering of the voids in the $1\,746$ Hydra ``wall
galaxies'' (the slope $-3/0.4$ is marked by the straight line).}
\label{Zipf-H}
\end{figure}

We now apply the void-finder to the population of $1\,746$ 
``wall galaxies''.  We find and rank order the voids, which
are a juxtaposition of simplices, as shown in 
Fig.\ \ref{voids-H} (we only show from the third to the sixth rank 
void for ease of visualization).
The rank-ordering plot, displayed in Fig.\ \ref{Zipf-H}, 
has a wide left region corresponding
to the transition to homogeneity and a scaling region with
the right slope $-3/D = -7.5$.
We have tried various void-finder parameters
with similar results. The scaling region begins with the 300th void
(approximately), with size $0.0001$ (corresponding to linear size
$\sim 0.05$).

\section{Discussion}

We propose a new method for finding voids in the galaxy distribution
based on discrete geometry constructions, namely, Delaunay
and Voronoi tessellations. This method defines a void as a
juxtaposition of Delaunay simplices separated from other voids by
reasonable criteria, similar to separation criteria used in
other void-finders. This definiton assumes that the total space
occupied by the sample is to be filled with voids of irregular shape.
The size of voids in a scaling point distribution run from nearly 
the sample's size to about the minimal interparticle
distance.  A convenient representation of the size distribution is the
rank-ordering Zipf's plot, which is a power-law for a scaling point
distribution (Zipf's law).

We have demonstrated that Zipf's law holds for uniformly scaling point
distributions, that is, pure fractals (or monofractals).  However, it
holds at the most over a scaling range as large as the scaling range
defining the fractal (by the number-radius relation). This is shown by
our study of fractals with transition to homogeneity.

On the other hand, we have argued, following previous work in the
literature, that one must consider multifractal rather than
monofractal models of the galaxy distribution. A multifractal
distribution can be considered as a superposition of fractals with
different dimensions. The simplest multifractal is a superposition of
two fractals, which is in accord with the usual division of galaxies
between wall and field galaxies.
Therefore, we have assumed that a multifractal galaxy distribution 
can be approximately described as
a superposition of just two fractal populations. So we have
built two galaxy populations from 
the halo distribution found in an $N$-body simulation 
(by the Hydra Consortium).
The strongly clustered population scales with a
very low fractal dimension ($D=0.4$), much smaller than the fractal
dimensions of the full galaxy distribution reported in the literature, 
and it has a transition to homogeneity. 
We have used our method to find the voids in this strongly clustered
population. Of course, these voids contain {\em inside} members of the
weakly clustered population.  The Zipf law holds for the voids, but
in a limited range.

A relatively small range of the Zipf law means that most sizeable
voids are outside the scaling region and are, therefore, strongly
influenced by the transition to homogeneity. This can help to explain
why no scaling is perceived in the void catalogues compiled in the
literature (Gaite \& Manrubia, 2002).  Even if the void-finder
employed is adequate, a poor selection of the sample and/or of its
voids can spoil scaling. Regarding the sample, one has to select a
sufficiently uniform population to be considered a monofractal. One
should check that the ``wall builder phase'' of the usual void-finding
procedure accomplishes it. Regarding the voids, the first hundreds
of voids may only show the transition to homogeneity of a scaling
distribution on smaller scales. However, it is usual to restrict the
search for voids to a few hundreds. Hopefully, improvement along these
lines will eventually show scaling in the void catalogues.



\begin{acknowledgments}
I thank Francisco Prada for a conversation and for pointing out the
work of Gottloeber et al (2003). My work is supported by a ``Ramon y
Cajal'' contract and by grant BFM2002-01014 of the Ministerio de
Ciencia y Tecnologia.
\end{acknowledgments}



%


\end{document}